\PassOptionsToPackage{unicode}{hyperref}
\PassOptionsToPackage{hyphens}{url}
\PassOptionsToPackage{dvipsnames,svgnames,x11names}{xcolor}
\documentclass[
  12pt]{article}

\usepackage{amsmath,amssymb}
\usepackage{iftex}
\ifPDFTeX
  \usepackage[T1]{fontenc}
  \usepackage[utf8]{inputenc}
  \usepackage{textcomp} 
\else 
  \usepackage{unicode-math}
  \defaultfontfeatures{Scale=MatchLowercase}
  \defaultfontfeatures[\rmfamily]{Ligatures=TeX,Scale=1}
\fi
\usepackage{lmodern}
\ifPDFTeX\else  
\fi
\IfFileExists{upquote.sty}{\usepackage{upquote}}{}
\IfFileExists{microtype.sty}{
  \usepackage[]{microtype}
  \UseMicrotypeSet[protrusion]{basicmath} 
}{}
\makeatletter
\@ifundefined{KOMAClassName}{
  \IfFileExists{parskip.sty}{%
    \usepackage{parskip}
  }{
    \setlength{\parindent}{0pt}
    \setlength{\parskip}{6pt plus 2pt minus 1pt}}
}{
  \KOMAoptions{parskip=half}}
\makeatother
\usepackage{xcolor}
\makeatletter
\ifx\paragraph\undefined\else
  \let\oldparagraph\paragraph
  \renewcommand{\paragraph}{
    \@ifstar
      \xxxParagraphStar
      \xxxParagraphNoStar
  }
  \newcommand{\xxxParagraphStar}[1]{\oldparagraph*{#1}\mbox{}}
  \newcommand{\xxxParagraphNoStar}[1]{\oldparagraph{#1}\mbox{}}
\fi
\ifx\subparagraph\undefined\else
  \let\oldsubparagraph\subparagraph
  \renewcommand{\subparagraph}{
    \@ifstar
      \xxxSubParagraphStar
      \xxxSubParagraphNoStar
  }
  \newcommand{\xxxSubParagraphStar}[1]{\oldsubparagraph*{#1}\mbox{}}
  \newcommand{\xxxSubParagraphNoStar}[1]{\oldsubparagraph{#1}\mbox{}}
\fi
\makeatother

\usepackage{longtable,booktabs,array}
\usepackage{tabularx}
\usepackage{calc} 
\usepackage{etoolbox}
\makeatletter
\patchcmd\longtable{\par}{\if@noskipsec\mbox{}\fi\par}{}{}
\makeatother
\IfFileExists{footnotehyper.sty}{\usepackage{footnotehyper}}{\usepackage{footnote}}
\makesavenoteenv{longtable}
\usepackage{graphicx}
\usepackage{pdflscape}
\makeatletter
\def\maxwidth{\ifdim\Gin@nat@width>\linewidth\linewidth\else\Gin@nat@width\fi}
\def\maxheight{\ifdim\Gin@nat@height>\textheight\textheight\else\Gin@nat@height\fi}
\makeatother
\setkeys{Gin}{width=\maxwidth,height=\maxheight,keepaspectratio}
\makeatletter
\def\fps@figure{htbp}
\makeatother

\makeatletter
\@ifpackageloaded{caption}{}{\usepackage{caption}}
\AtBeginDocument{%
\ifdefined\contentsname
  \renewcommand*\contentsname{Table of contents}
\else
  \newcommand\contentsname{Table of contents}
\fi
\ifdefined\listfigurename
  \renewcommand*\listfigurename{List of Figures}
\else
  \newcommand\listfigurename{List of Figures}
\fi
\ifdefined\listtablename
  \renewcommand*\listtablename{List of Tables}
\else
  \newcommand\listtablename{List of Tables}
\fi
\ifdefined\figurename
  \renewcommand*\figurename{Figure}
\else
  \newcommand\figurename{Figure}
\fi
\ifdefined\tablename
  \renewcommand*\tablename{Table}
\else
  \newcommand\tablename{Table}
\fi
}
\@ifpackageloaded{float}{}{\usepackage{float}}
\floatstyle{ruled}
\@ifundefined{c@chapter}{\newfloat{codelisting}{h}{lop}}{\newfloat{codelisting}{h}{lop}[chapter]}
\floatname{codelisting}{Listing}

\makeatother
\makeatletter
\@ifpackageloaded{caption}{}{\usepackage{caption}}
\@ifpackageloaded{subcaption}{}{\usepackage{subcaption}}
\makeatother

\ifLuaTeX
  \usepackage{selnolig}  
\fi
\usepackage[]{natbib}
\usepackage{bookmark}

\IfFileExists{xurl.sty}{\usepackage{xurl}}{} 
\hypersetup{
  pdftitle={Title},
  pdfauthor={Emanuela Furfaro; Katherine Elizabeth Lovelace; Yuhan Zhang; Simone Mosciatti},
  pdfkeywords={3 to 6 keywords, that do not appear in the title},
  colorlinks=true,
  linkcolor={blue},
  filecolor={Maroon},
  citecolor={Blue},
  urlcolor={Blue},
  pdfcreator={LaTeX via pandoc}}

\usepackage{multicol} 

\usepackage[most]{tcolorbox}
\newtcolorbox{conversation}{breakable, enhanced, colback=gray!5, colframe=gray!60, boxrule=0.5pt, arc=2pt, left=6pt, right=6pt, top=4pt, bottom=4pt, fontupper=\footnotesize, before upper={\setlength{\parskip}{3pt}}}

\newcommand{\anon}{1}

\begin{document}

\def\spacingset#1{\renewcommand{\baselinestretch}%
{#1}\small\normalsize} \spacingset{1}


\if1\anon
{
  \title{\bf 
   AI-Guided Assignments in Practice:\\ Students' Experiences in a Large Introductory Statistics Course}
  \author{Emanuela Furfaro\thanks{Corresponding author: \href{mailto:efurfaro@uw.edu}{efurfaro@uw.edu}}\,$^{1}$, Katherine Elizabeth Lovelace$^{1}$, Yuhan Zhang$^{2}$,\\ Simone Mosciatti$^{3}$ \\[1em]
    \normalsize $^{1}$Department of Statistics, University of Washington, Seattle, WA, USA \\
    \normalsize $^{2}$Department of Biostatistics, University of Washington, Seattle, WA, USA \\
    \normalsize $^{3}$Independent Researcher}
\date{}
  \maketitle
} \fi

\if0\anon
{
  \bigskip
  \bigskip
  \bigskip
  \begin{center}
    {\LARGE\bf Title}
\end{center}
  \medskip
} \fi

\bigskip
\begin{abstract}
Large language models (LLMs) raise new questions for instructors, among them how to integrate LLMs productively into coursework and what to evaluate in students' work when an LLM can produce much of it directly.
LLM-guided homework assignments offer one answer.
Instructors configure the LLM to guide students in ways that achieve specific pedagogical goals, such as recalling material, discovering new concepts, or critically evaluating deliberately flawed reasoning, and grade the full student--LLM conversation, and with it the student's reasoning, rather than only the final answer.
In this paper, we report on the large-scale deployment of such assignments, implemented through LLteacher, an open-source tool that supports recall, discovery, and critical-thinking assignments, in a large introductory statistics course serving mostly pre-major undergraduates.
Results on 134 students show meaningful gains in self-reported confidence, with over 80\% agreeing that the assignments improved their comprehension of course material and their ability to interpret statistical results; gains were largest among students who started with lower confidence, suggesting this approach may help narrow gaps in preparation.
Interaction patterns varied widely, from long single messages to rapid iterative exchanges, and hallucinations, when present, were rarely disruptive to learning.
\end{abstract}

\noindent%
{\it Keywords:} technology innovations, educational techniques, generative AI, large language models, homework\\

\vfill

\newpage
\spacingset{1.8} 

\section{Introduction}

The emergence of large language models (LLMs) has reshaped how students interact with educational content and it confronts instructors with two challenges: how to integrate LLMs into classwork productively, and what to evaluate in students' work since an LLM can produce much of it directly. 
When used without pedagogical structure, systematic reviews find that the use of Generative AI produces mixed or null learning benefits \citep{hon-2026}. Studies of unstructured AI use in STEM courses document patterns of dependency, uncritical trust in incorrect outputs, and passive solution-seeking rather than active problem-solving \citep{wang-et-al-2025}. 
Critically, these harms are not distributed equally: students with stronger academic backgrounds are better positioned to evaluate and productively use AI-generated content, raising concerns that unguided LLM access may amplify existing educational inequities \citep{wang-et-al-2025, baek-et-al-2024}. 
Educators must therefore design strategies that integrate AI in ways that preserve active engagement, support critical thinking, and promote equitable access.
In addition, traditional homework evaluation focuses on the final output, a target LLMs can now produce directly, which does little to encourage the productive struggle that homework is meant to build. 
Addressing this requires tools designed to make the reasoning process itself visible, structured, and gradable, rather than tools that only restrict or monitor LLM outputs.


To address these challenges in the specific context of homework assignments in Statistics, \cite{furfaro-mosciatti-2026} developed LLteacher, an open-source web application that operationalizes a structured approach to AI-assisted learning. LLteacher guides students through iterative problem-solving, preserving the cognitive processes that homework is designed to reinforce. 
Importantly, LLteacher does not only constrain how students use AI, it also expands how instructors can design assignments, enabling novel pedagogical goals that would be difficult to implement without LLM support, and it allows instructors to access the students' reasoning process, to evaluate not only the final output but also the students' thinking \citep{furfaro-mosciatti-2026}.

In this paper, we report on the use of LLM-guided homework assignments, implemented through LLteacher, in a large introductory statistics course serving pre-major undergraduates.
Specifically, we make three contributions: (1) we show that LLM-guided assignments can be successfully deployed in a large classroom setting, establishing their feasibility as a practical instructional approach; (2) using data from 134 students, we provide preliminary evidence on their efficacy, using pre- and post-assignment self-assessments to track perceived learning gains; and (3) we report students' perceptions of and experiences with this kind of assignment.
Together, these results contribute to the literature on the integration of structured AI assistance into statistics courses, documenting how LLM-guided assignments can be effectively deployed, what effect they have on student learning, and how students respond to them.

\section{Background and Related Work}

The integration of AI into education predates the era of LLMs.
Traditional Intelligent Tutoring Systems (ITS) used hand-authored knowledge
representations, such as production rules grounded in Anderson's ACT-R theory
\citep{anderson1996act} or constraint-based models \citep{ohlsson1996learning}, to
generate targeted feedback by comparing student responses to predefined models of correct
reasoning. While effective in several domains, including Statistics
\citep{hobert-berens-2024}, these systems required substantial expert effort to build and did not scale easily to new content. The availability of LLMs has opened a different path: adaptive, natural-language feedback can now be generated without hand-authored rules or large interaction datasets \citep{stamper2024enhancing},
dramatically lowering the barrier to deploying AI-assisted learning in new courses.

This has prompted a wave of research examining the educational effects of LLMs in STEM disciplines, with results that are mixed.
On the positive side, several studies in statistics suggest that LLMs can support conceptual understanding and increase student
engagement \citep{hariyanti-et-al-2025}. \cite{wahba-et-al-2024} conducted a
quasi-experimental study comparing traditional instruction to ChatGPT-assisted learning in a probability and statistics course, finding improvements in both statistical reasoning and student attitudes in the AI-assisted group.
\cite{bien-mukherjee-2025} successfully
integrated GitHub Copilot into an introductory statistics course, where students learned to write English prompts that generated executable R code. \cite{kula-2026} conducted a study that examined how students in a second-year introductory statistics course used a restricted LLM, Alexandria. They saw that students were not just looking for the final answer but were using the LLM as a tool for their learning. They found that students use the LLM for three main purposes: the first was student procedural execution (e.g. calculating a CI), the second was conceptual clarification (e.g., Type I vs. Type II error), and the third was iterative reinforcement of statistical understanding (e.g. returning to conversations and checking final answers), showing that AI and LLM integration into the statistical pedagogy is a helpful tool and resource. 

However, not all studies have depicted a fully positive picture. 
\cite{bastani-et-al-2024}, in a high school mathematics context, found that students who used GPT-4 initially outperformed peers but subsequently performed worse than students who never had access once the tool was removed, a pattern suggestive of dependency rather than durable learning.
In an introductory physics course, students expressed high trust in ChatGPT's responses even when they were incorrect \citep{ding-et-al-2023}. \cite{wang-et-al-2025} found that over half of STEM undergraduates primarily used ChatGPT to obtain direct solutions with limited active problem-solving, and that usage patterns varied substantially across demographic groups, raising concerns about reinforcing existing inequities.  These findings are consistent with a recent systematic review concluding that the learning benefits of generative AI in higher education are sensitive to implementation design, with mixed or null effects when AI is used without pedagogical structure \citep{hon-2026}.  Together, this evidence suggests that the educational value of generative AI depends critically on \emph{how} it is integrated into learning activities.

In response, researchers in STEM education have developed tools that constrain how LLMs respond to students.
CodeAid \citep{kazemitabaar2024codeaid}, deployed in a programming course of approximately 700 students, prevented the LLM from providing direct
code solutions; students reported that it promoted learning better than general-purpose tools like ChatGPT by guiding them to reason through problems rather than handing over answers.
LPITutor \citep{liu2025lpitutor} combines retrieval-augmented generation (RAG)
with adaptive prompt engineering to ground LLM responses in course-specific documents, achieving high factual accuracy and user satisfaction.
More recently, \cite{sunil-thakkar-2025} introduced SocraticAI, a tutoring system for undergraduate computer science. Students must ask well-formed questions, their daily use is limited, and the system guides each interaction through Socratic dialogue. 
In a pilot deployment, over 75\% of students progressed from vague help-seeking to sophisticated problem decomposition within two to three weeks.
\cite{happe-et-al-2025} report on RockStartIT, an AI-powered assistant developed for digital programming and computational thinking courses.
The results on 13 students and instructors show that learners valued the low-stakes environment and the scaffolded guidance it provided, reporting high degrees of perceived usefulness.

While most of these tools have been developed and evaluated in computer science courses, the educational tools and associated evidence base in statistics is considerably thinner. 
QuickTA, a large language model-based teaching assistant, was introduced by \cite{kumar-et-al-2023} to explore how LLMs could support students at scale. At the University of Toronto, \cite{oh-et-al-2026} later used QuickTA in a cluster-randomized study comparing a trained vs. untrained Llama 3.1-based version of the tool in a second-year introductory statistics course, testing whether the trained version improved graded tutorial performance. They found no significant difference in student gains, and two thirds of students rated QuickTA as unhelpful or neutral, though students with lower final scores and international students found it useful. \cite{amanlou-et-al-2026} examined guided LLM use in an undergraduate Probability and Statistics course, finding that, compared with unrestricted access, guided use was associated with more learning-oriented interaction patterns, including prioritizing reasoning over final answers and requesting stepwise support.

At Northwestern's Kellogg School of Management, \cite{bray2024} observed that most of their MBA data analysis coursework could already be completed with ChatGPT, and in response developed Data Science with Large Language Models, a course that fully integrates generative AI into student coursework rather than treating it as a supplementary tool. \cite{braymatrin2025} extended this work with an RCT comparing a custom-prompted GPT-4o to the base model across assignments, hypothesizing that purpose-built AI agents would improve student satisfaction, effort, and learning. 
They found evidence only for greater enjoyment: student gains and performance did not differ significantly between groups.

In this paper, we use LLteacher, an open-source LLM-based tool developed by \cite{furfaro-mosciatti-2026} in the context of statistics courses.
Unlike the systems discussed above, which adopt a single pedagogical approach such as Socratic questioning or hint-based scaffolding, LLteacher supports multiple instructor-configurable modes to create assignments that achieve different pedagogical goals as discussed in Section~\ref{sec:design}, allowing instructors to configure distinct learning objectives across different assignments. 
This feature also allows instructors to not simply use LLMs in their teaching, but to leverage them to redesign their assignments and rethink their goals.
In addition, LLteacher surfaces student–AI interaction data, allowing instructors to grade the whole interaction and the reasoning process, rather than just the final result.
Another distinguishing feature is that LLteacher allows instructors to redesign assignments to integrate LLMs as an integral part of the course, similarly to the approach in \cite{bray2024}, rather than using LLMs tools as an additional resource for learning.

The present paper reports the first large-scale deployment of LLteacher in an introductory statistics course. 
We characterize student interaction patterns, report student perceptions of the tool, and assess the extent to which the system supported intended learning outcomes. 
Because LLteacher is open-source, this deployment is fully replicable, and the tool is available for adoption by instructors in analogous settings.

\section{LLteacher: Design and Functionality}
\label{sec:design}

LLteacher is a web-based application built in Python using the Django framework \citep{furfaro-mosciatti-2026}. 
LLteacher employs SQLite for data storage and integrates with several LLMs, including models from OpenAI (e.g., GPT-5), Google (Gemini), and Anthropic (Claude), via API. 
LLteacher is based on a two-prompt engineering strategy which allows instructors to customize LLteacher's behavior to achieve different pedagogical goals (level 1 prompt) and to provide solutions to the assignments to ground LLteacher's responses \citep{furfaro-mosciatti-2026}.
LLteacher's code is fully open-source (MIT license), with version 1.0.0 archived on Zenodo \citep{mosciatti-furfaro-2026-code}, enabling customization and extension to other courses or LLMs.

The platform consists of two main interfaces:

\begin{itemize}

\item Instructor View: Educators can create and manage assignments by specifying problem statements and reference solutions (level 2 prompt). 
This view also provides configuration options, allowing instructors to select the underlying LLM, customize prompts, and adjust system behavior to align with assignment's objectives (level 1 prompt). Instructors can review the complete student–LLM interactions, allowing instructors to grade the quality of the interaction along with the students' final answer, and creating opportunities for formative assessment and personalized feedback.

\item Student View: Students can see the available assignments along with the due date. 
They click on the assignment they want to complete and start a conversation with LLteacher.
LLteacher acts as a coach rather than a solver: it poses questions, offers hints, and encourages problem-solving without revealing full solutions.
\end{itemize}

We used LLteacher to design three types of assignments aimed at achieving:

\begin{itemize}
\item \textbf{Recall-based learning}: reinforces material already taught by prompting students to organize their thoughts, check assumptions, and articulate reasoning steps (e.g., ``What assumptions do we need to check before this test?''), strengthening retention through retrieval practice.
\item \textbf{Discovery-based learning}: guides students to construct new knowledge themselves through a sequenced progression of questions and hints (e.g., sampling repeatedly to build intuition for confidence intervals) before formal instruction is given.
\item \textbf{Critical thinking (error introduction)}: the LLM deliberately gives a flawed definition or step, and students must catch, explain, and correct the error, building skepticism toward AI outputs and reinforcing conceptual understanding.
\end{itemize}

These goals are embedded in the way the instructor configures each assignment.
Underlying all approaches is error-driven learning, a principle that views mistakes as opportunities for growth rather than failures \citep{tulis-et-al-2016}, which is difficult to achieve in traditional assignments.

In addition to interactive assignment types, LLteacher supports non-interactive question formats and embeddable response widgets, which were used in this study to administer pre- and post-assessment items at the start and end of each assignment (described further in Section~\ref{sec:data}).

\section{In-Class Deployment and Data}
\label{sec:data}

\subsection{Class Context}

LLteacher was used during the Spring 2026 quarter in an introductory statistics course serving undergraduate pre-majors with some prior exposure to calculus.
The course spans descriptive statistics to an introduction to probability, inference and linear regression.
The results presented in this paper refer to 134 students who gave consent for their data to be used for research purposes.
The study was reviewed and approved by the University of Washington Human Subjects Division (IRB ID: STUDY00025402).

The course structure included traditional homework assignments, laboratory sessions led by the teaching assistants, in-class written examinations and six LLteacher assignments.
Students were allotted five days to complete each assignment, with assignments released approximately weekly, starting in week 4.

\subsection{Assignment Design}

Each released LLteacher assignment embedded one or more of the pedagogical goals described in Section~\ref{sec:design} within a topic drawn from the course syllabus. Table~\ref{tab:assignment_summary} summarizes the topic, pedagogical goal(s), and task for each assignment.
For all assignments, we used Claude Sonnet 4 (Anthropic) as the underlying LLM.
Topics increased in difficulty as the quarter progressed. The earliest assignments were deliberately straightforward, allowing students to become accustomed to working with LLteacher, while later assignments introduced critical-thinking and discovery components that placed increasing demands on students.

\begin{table}[!h]
\centering
\small
\begin{tabularx}{\textwidth}{c l X}
\toprule
Assignment & Mode(s) & Topic \\
\midrule
1 & Recall & Expectation and variance of linear combinations of random variables (stock portfolio) \\
2 & Recall & Normal approximation to the binomial \\
3 & Recall + Discovery & Confidence intervals (two-sided $\to$ one-sided) \\
4 & Recall + Critical thinking & ANOVA (variability explained vs.\ unexplained; $F$-test $p$-value) \\
5 & Recall + Critical thinking & Checking linear regression assumptions \\
6 & Discovery & Multiple linear regression (extra credit assignment) \\
\bottomrule
\end{tabularx}
\caption{Summary of assignment topics and pedagogical modes.}
\label{tab:assignment_summary}
\end{table}

To illustrate how these modes were operationalized in practice, and to contrast the beginning and end of the quarter, we present two examples: Assignment 1 (Recall) and Assignment 6 (Discovery).

\begin{figure}[H]
    \centering
    \begin{subfigure}{0.6\textwidth}
        \centering
        \includegraphics[width=\linewidth]{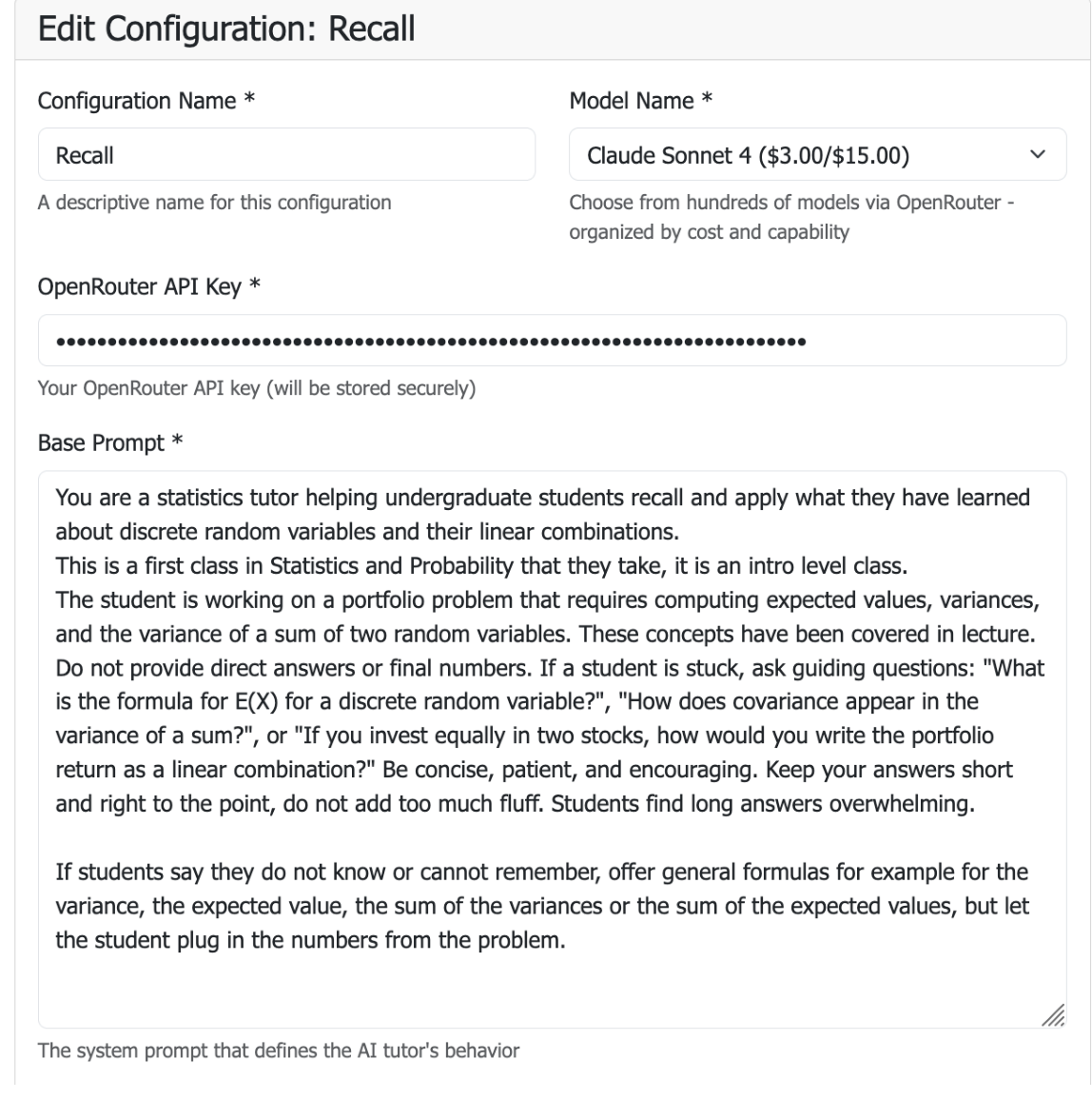}
        \caption{LLM base prompt.}
        \label{fig:a1_config}
    \end{subfigure}

    \vspace{0.5em}
    \begin{subfigure}{\textwidth}
        \centering
        \includegraphics[width=\linewidth]{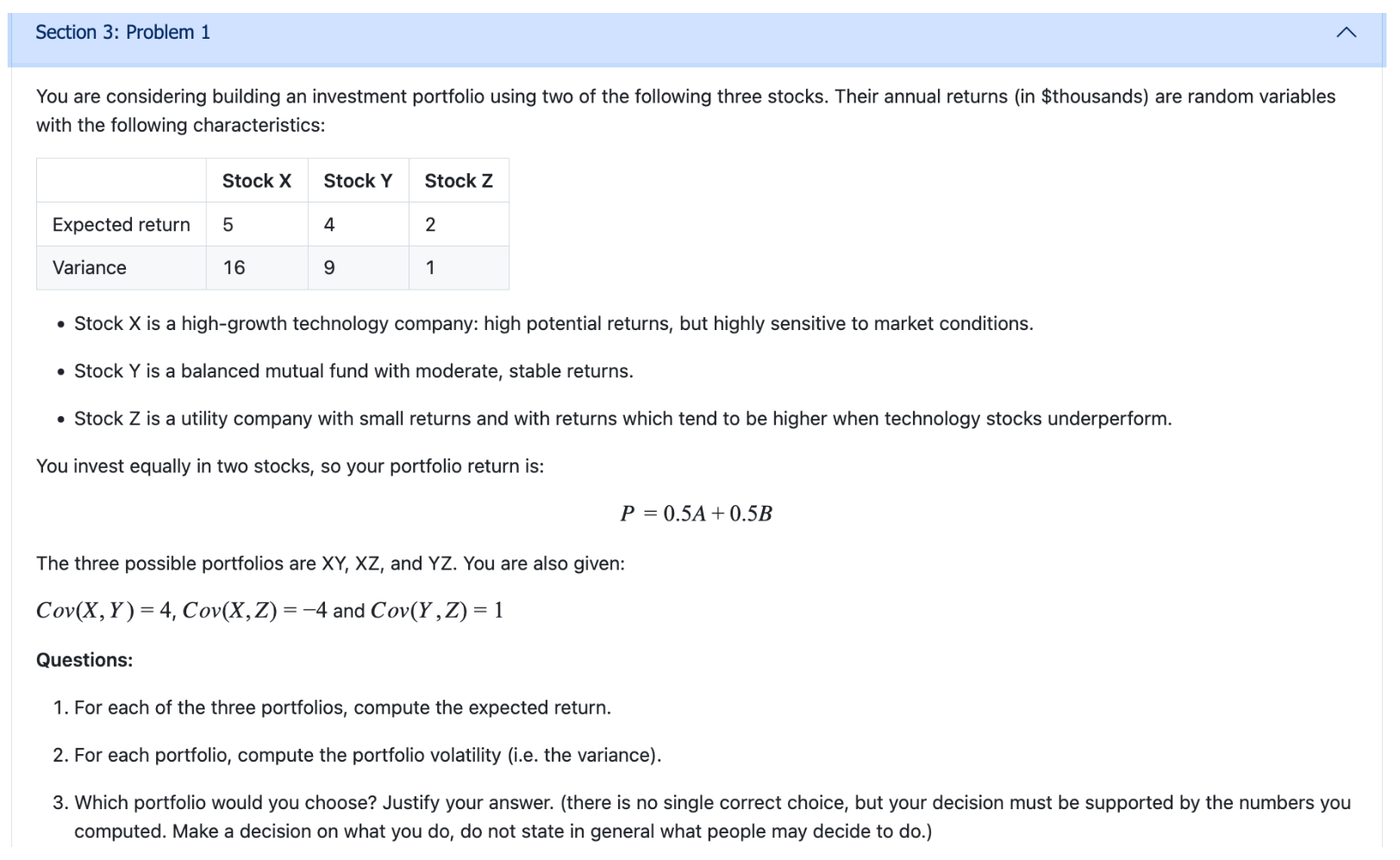}
        \caption{Problem text as presented to students (Problem 1).}
        \label{fig:a1_text}
    \end{subfigure}
    \caption{Assignment 1 (Recall): LLM base prompt (a) and problem text (b).}
    \label{fig:a1}
\end{figure}

In Assignment 1, students worked through a portfolio problem requiring them to compute the expected return and variance of linear combinations of stock returns, with given covariances. The LLM was instructed to prompt students toward the relevant formulas without supplying them outright, asking guiding questions such as ``how does covariance contribute in the variance of a sum of variables?''. If students could not recall a formula, the LLM was to provide its general form (e.g., the variance of a sum) while requiring students to substitute the problem's own numbers. 
Figure~\ref{fig:a1} shows the LLM base prompt configured for this assignment (Figure~\ref{fig:a1_config}) and the corresponding problem text presented to students (Figure~\ref{fig:a1_text}).

The excerpt below, drawn from an anonymized student conversation on Assignment 1, illustrates four moments of this scaffolding in practice: the student asking clarifying questions, the LLM prompting the student to recall a formula rather than supplying it unprompted, catching a mislabeled quantity mid-calculation, and validating the student's open-ended justification in Question 3 without dictating which portfolio was ``correct.''

\begin{conversation}

\noindent\textbf{Student:} What does expected return mean in the context of this problem?

\noindent\textbf{LLteacher:} Great question! In this problem, the expected return is the average return you can expect from a stock over time. [\ldots] It's the same concept as the expected value $E(X)$ of a random variable: the long-run average outcome. [\ldots] Do you remember the formula for the expected value of a linear combination of random variables?

\noindent\textbf{Student:} I do not, could you please refresh me.

\noindent\textbf{LLteacher:} No problem! Here's the key formula: $E(aX + bY) = aE(X) + bE(Y)$[\ldots] Now can you try calculating the expected return for the XY portfolio using the given expected returns?

\hfill [\ldots]

\noindent\textbf{Student:} Yes; I would get $(0.5)^2 \times 16 + (0.5)^2 \times 9 + 2\times(0.5)(0.5)\times 4 = Cov(X,Y)$.

\noindent\textbf{LLteacher:} You've set up the formula correctly! But I think there's a small mix-up at the end. [\ldots] this should equal $Var(P_{XY})$, not $Cov(X,Y)$. The 4 in your calculation \emph{is} the $Cov(X,Y)$ that you're plugging into the formula. [\ldots] Can you calculate what this equals numerically?

\noindent\textbf{Student:} Ah yes, thanks for catching my mistake. This equates to 8.25 numerically.

\hfill [\ldots]

\noindent\textbf{LLteacher:} For question 3, which portfolio would you choose and why? Think about the trade-off between expected return (higher is better) and variance/risk (lower is better).

\noindent\textbf{Student:} I would choose the XZ portfolio, as it is not as volatile as the XY portfolio, which means it is safer, but not as low as the YZ portfolio, which has a very low return.

\noindent\textbf{LLteacher:} Excellent reasoning! You've made a well-supported decision based on the numbers you calculated. [\ldots] You correctly identified the key trade-off between risk and return, and chose a portfolio that balances both factors rather than going to either extreme.

\end{conversation}

\begin{figure}[H]
    \centering
    \includegraphics[width=0.7\textwidth]{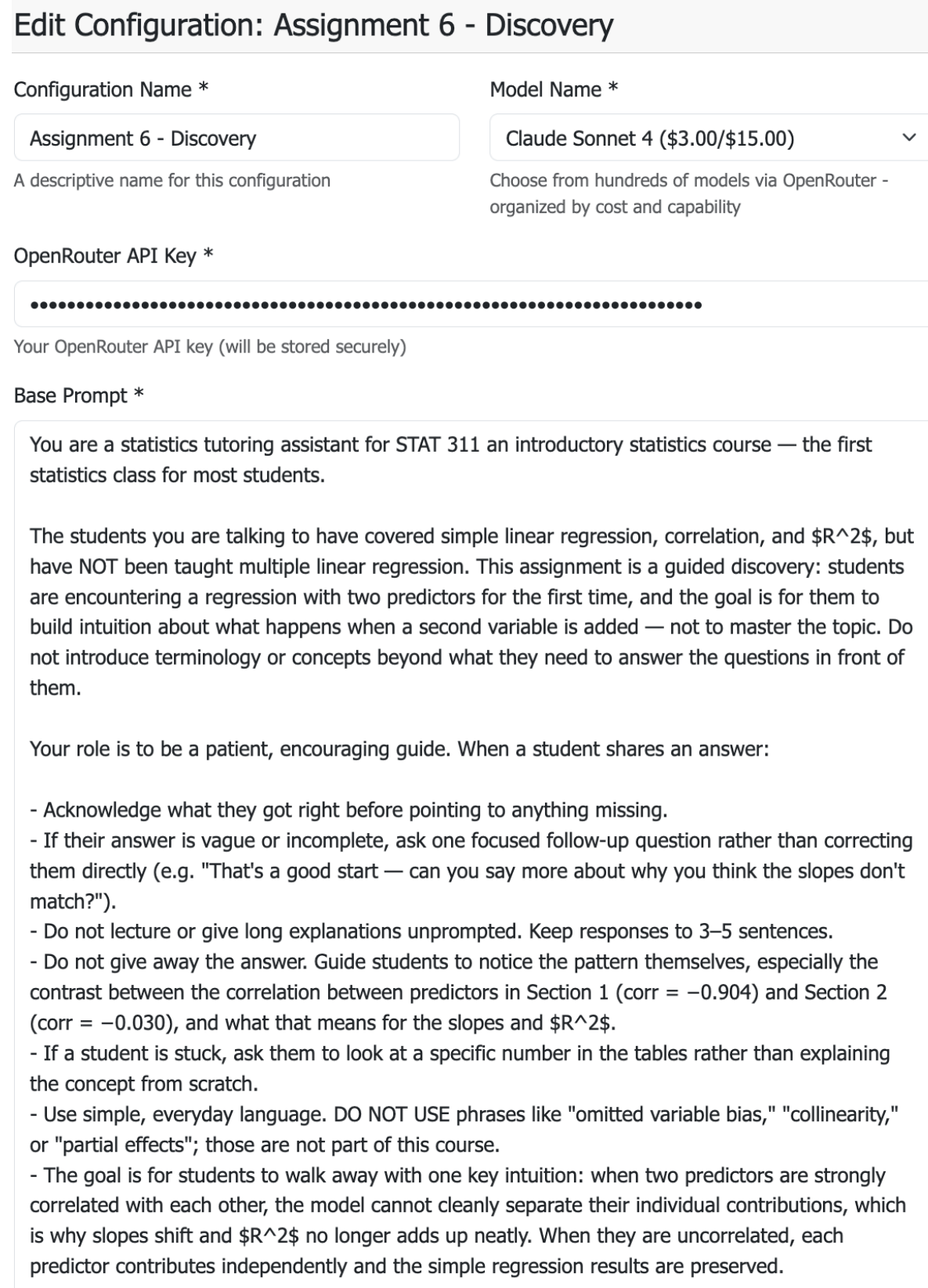}
    \caption{LLM base prompt for Assignment 6 (Discovery).}
    \label{fig:a6_config}
\end{figure}

\begin{figure}[H]
    \centering
    \includegraphics[width=\textwidth]{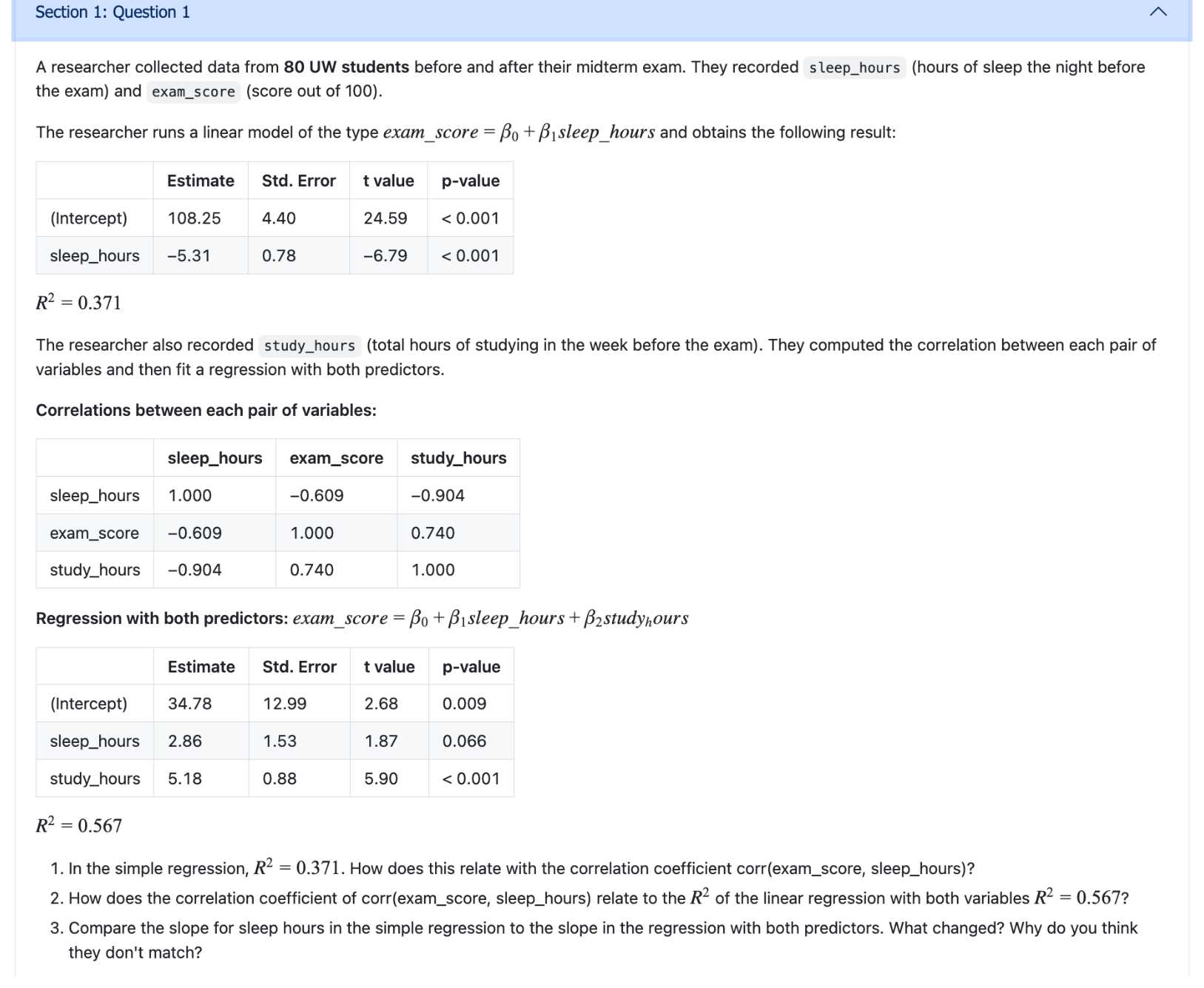}
    \caption{Problem text as presented to students: Assignment 6, Question 1.}
    \label{fig:a6_text}
\end{figure}

In Assignment 6, students fit a regression of exam score on sleep hours, then added a second predictor and observed that the slope and $R^2$ shifted substantially in one case but barely at all in another. Rather than explaining the technicalities, the LLM was instructed to guide students toward noticing how, when going from a simple linear regression to a multiple linear regression, the correlation between predictors affects the coefficients. 
Figure~\ref{fig:a6_config} shows the LLM base prompt configured for this assignment, and Figure~\ref{fig:a6_text} shows the corresponding problem text presented to students.

The excerpt below, drawn from an anonymized student conversation on Assignment 6, illustrates two moments of this guided discovery: the LLM redirecting the student to the correct correlation coefficient rather than confirming a misreading of the table, and the student arriving, through the LLM's questioning, at the reason why the slope for sleep hours flips sign once study hours is added to the model.

\begin{conversation}

\noindent\textbf{Student:} The r in the table gives 0.78, which is a stronger positive relationship between the two variables.

\noindent\textbf{LLteacher:} I think you might be looking at the wrong correlation. [\ldots] What is the correlation between exam score and sleep hours specifically? [\ldots] Also, pay attention to whether that correlation is positive or negative. Does it have a minus sign in front of it?

\noindent\textbf{Student:} Oops, it is $-0.609$, which is a decent negative relationship between the two.

\hfill [\ldots]

\noindent\textbf{LLteacher:} Look at the slope for sleep hours in the simple regression ($-5.31$) versus the slope in the two-predictor model ($+2.86$). What's the biggest change you notice?

\noindent\textbf{Student:} I notice that in the single-variable regression, the slope is negative and large, whereas in the two-variable regression, the slopes are both positive.

\noindent\textbf{LLteacher:} Great observation! [\ldots] Why do you think this happened? Look at the correlation table. What do you notice about the relationship between sleep hours and study hours? How might that strong relationship between the two predictors help explain why the slope for sleep hours changed so dramatically?

\hfill [\ldots]

\noindent\textbf{Student:} If someone sleeps a lot, it means they probably study less, which is why the first line is so strongly negative.

\noindent\textbf{LLteacher:} Exactly! You've got it! [\ldots] Once we add study hours to the model, [\ldots] the model can see the benefit of sleep and the benefit of studying independently. That's why sleep hours gets a positive slope once study hours is included. Does this make sense?

\noindent\textbf{Student:} Yes, this makes sense! Thank you.

\end{conversation}

\subsection{Dataset 1: Interaction Logs and Pre/Post Self-Ratings}

All 134 consenting students submitted at least two LLteacher assignments, with 90\% of them completing all six assignments. 
At the beginning of each assignment, students were asked to self-rate their confidence in the topic to be covered; typically via two items: Widget 1, addressing mathematical reasoning and understanding (the underlying mathematical or logical reasoning behind the relevant statistical procedure), and Widget 2, addressing interpretation of the procedure's results in context.
Upon completing the assignment, students were asked to rate themselves again on the same two items. 
This pre/post design was initially implemented using open-ended prompts; beginning with Assignment 3, however, the assessment was administered through a structured widget with a sliding scale from 0 to 10 that displayed each student's initial response alongside the post-assignment prompt, reducing recall bias.

Student interactions with LLteacher are logged automatically as part of the system's architecture and were therefore available for analysis. For the purposes of this paper, interaction data were summarized at the student-assignment level using two metrics: the total number of messages exchanged and the student's average number of characters per message. 
Together, these measures provide a proxy for the depth and extent of each student's engagement with the tool.

\subsection{Dataset 2: End-of-Term Survey}

At the end of the quarter, students were invited to complete a survey assessing their experience with LLteacher. Among the 134 consenting students, 90\% completed the survey, whose questions are contained in Table~\ref{table-survey}.
The survey elicited student perceptions of LLteacher's utility in supporting their learning goals, their overall experience with the platform, and broader reflections on LLteacher's influence on their attitudes toward statistics and toward the use of AI as a learning tool.

\begin{table}[!h]
\centering
\footnotesize
\renewcommand{\arraystretch}{0.8}
\setlength{\tabcolsep}{4pt}
\begin{tabularx}{\textwidth}{c X l}
\toprule
Q\# & Item & Type \\
\midrule
1 & LLteacher increased my interest in Statistics & Rating (1--5) \\
2 & LLteacher helped me understand the concepts of this class & Rating (1--5) \\
3 & Using LLteacher improved my ability to interpret statistical analysis & Rating (1--5) \\
4 & Using LLteacher improved my learning in this class & Rating (1--5) \\
5 & LLteacher made me think more critically about AI-generated responses & Rating (1--5) \\
6 & LLteacher confused me & Rating (1--5) \\
7 & I found LLteacher useful & Rating (1--5) \\
8 & I would like to use LLteacher again & Rating (1--5) \\
9 & I enjoyed using LLteacher & Rating (1--5) \\
10 & LLteacher increased my interest in AI tools & Rating (1--5) \\
11 & I now use AI tools more & Rating (1--5) \\
12 & I am more confident in using AI tools for learning & Rating (1--5) \\
13 & I would like more classes to use AI tools & Rating (1--5) \\
14 & What aspects did you like the most about LLteacher? & Select all that apply \\
15 & What aspects did you like the least about LLteacher? & Select all that apply \\
16 & Did LLteacher hallucinate for you? & Multiple choice \\
17 & How problematic were these hallucinations for your learning or completion of the assignment? & Multiple choice \\
18 & Please share any additional feedback or comments you may have. & Short answer. \\
\bottomrule
\end{tabularx}
\caption{End-of-term survey items. Q1--Q13 use a five-point strongly-disagree-to-strongly-agree scale; Q6 is negatively worded.}
\label{table-survey}
\end{table}

\subsection{Grading}

Prior to the first assignment, students were given a short guide explaining what an LLteacher assignment is and how to interact productively with the tool. 
The guide framed each assignment as a guided conversation rather than a question-answering exercise, setting explicit expectations: engage actively with LLteacher's questions rather than waiting to be told the answer, show reasoning even when unsure, and answer the questions posed rather than skipping past them. 
It also modeled a set of good practices for interacting with LLMs, which included asking for a simpler explanation or a different example, asking to slow down, stating one's background so explanations could be calibrated, checking one's own understanding, pushing back when something seemed wrong, and summarizing a step before moving on. 
We also provided examples contrasting a poor conversation with a well-conducted one.

Grading followed the same three expectations set out in the guide. 
Because a correct numerical answer can be reached without any evidence of understanding, credit was not based on the final answer alone. 
Instead, graders looked for active engagement, whether the student read LLteacher's questions carefully and responded, and reasoning shown.

\section{Results}



\subsection{Description of Interactions}

\begin{figure}[!h]
\centering
\begin{minipage}{0.48\linewidth}
\centering
\includegraphics[width=\linewidth]{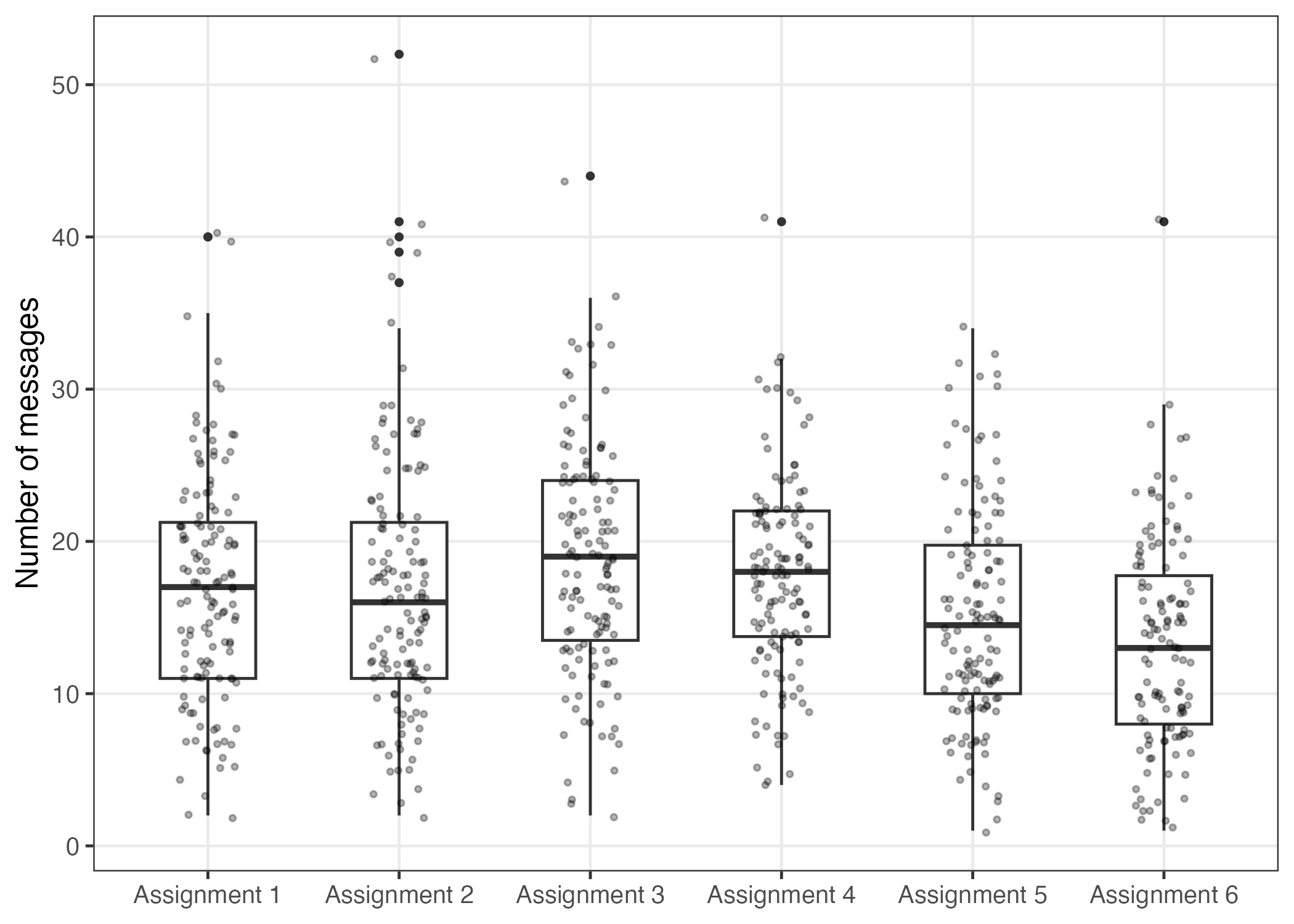}
\end{minipage}
\hfill
\begin{minipage}{0.48\linewidth}
\centering
\includegraphics[width=\linewidth]{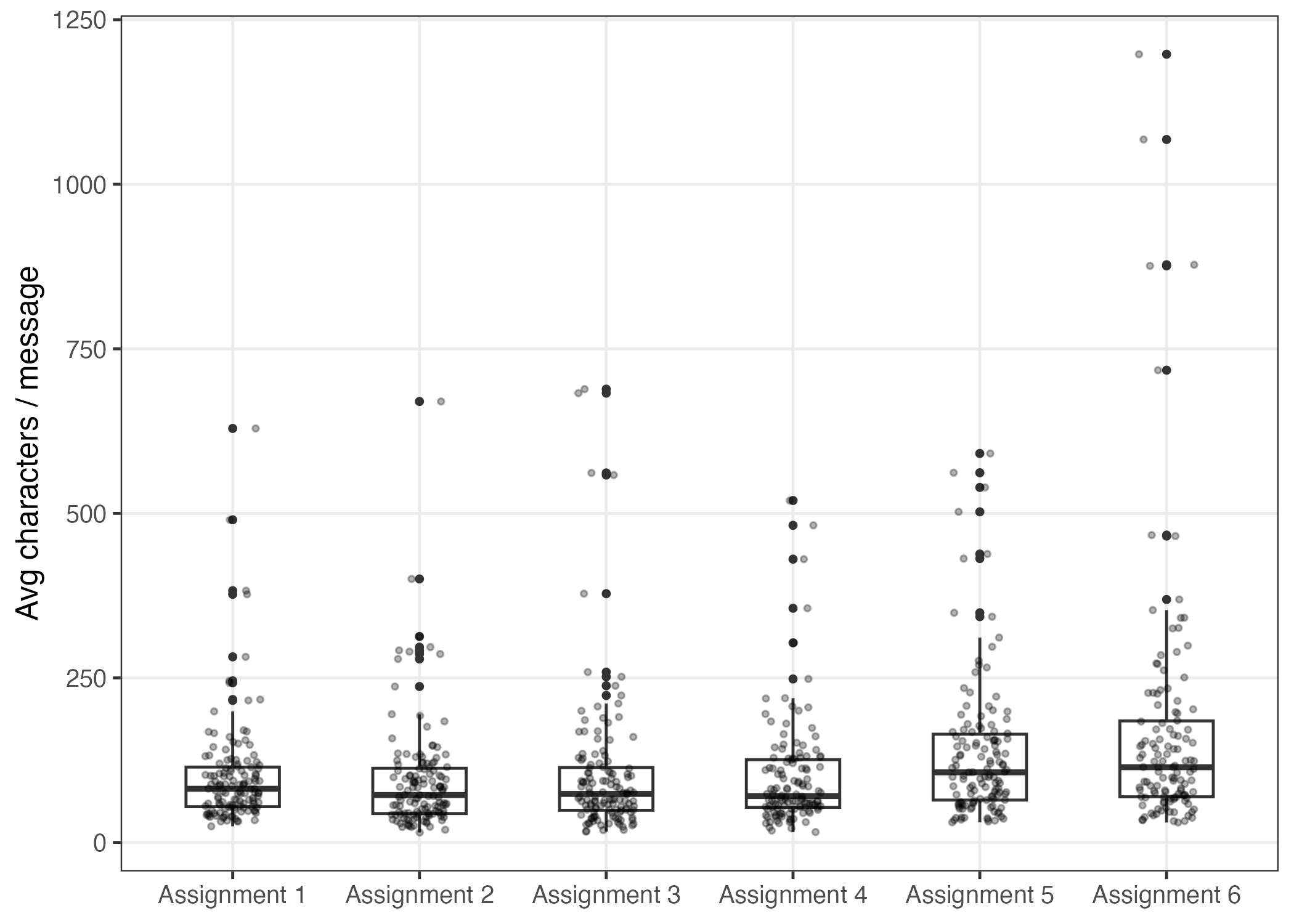}
\end{minipage}
\caption{Number of messages (left) and average characters per message (right) per student, by assignment. Boxplots show the median and interquartile range; points are individual students. \label{fig:interaction_by_assignment}}
\end{figure}

All students completed at least two assignments. The number of messages students sent remained roughly consistent across the six assignments, with a median around 15 messages (Figure~\ref{fig:interaction_by_assignment}, left). Assignment 5 saw somewhat fewer messages, as it contained only one question rather than two; Assignment 6, though it had two questions, was extra credit, so we would expect a different type of student to have completed it.

Message length was similarly stable (Figure~\ref{fig:interaction_by_assignment}, right), with most students sending messages of 50 to 100 characters, though the final assignment shows longer messages.

\begin{figure}[!h]
\centering
\includegraphics[width=0.85\linewidth]{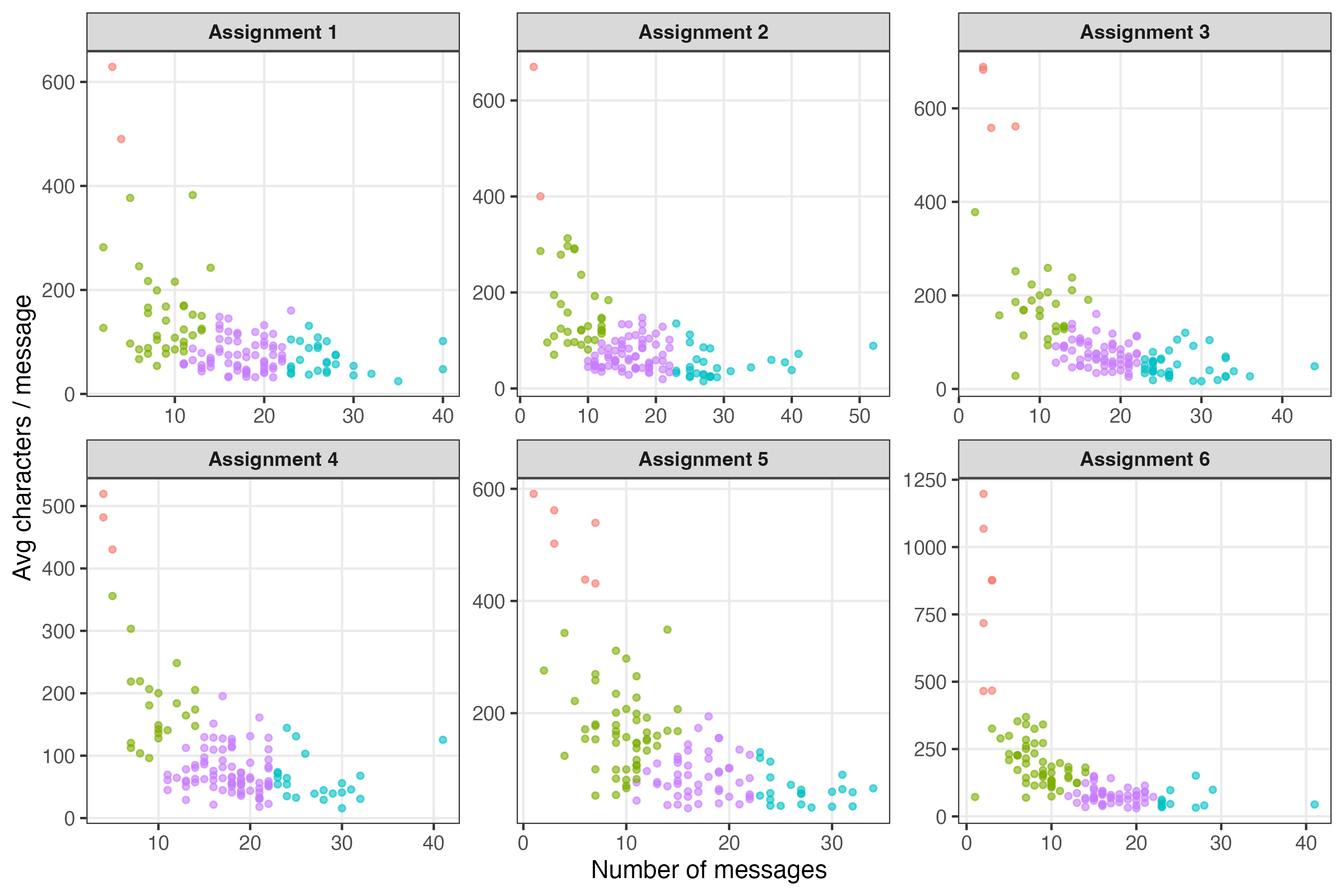}
\caption{Average characters per message against number of messages per student, faceted by assignment (Assignments 1 to 6). Each point is one student, colored by interaction style. \label{fig:avgchars_vs_nmsg_clusters}}
\end{figure}

Figure~\ref{fig:avgchars_vs_nmsg_clusters} shows the joint distribution of message length and message count, plotting average characters per message against number of messages for each student, and reveals four recurring interaction styles that appear consistently across all six assignments.
Using k-means clustering, we detected the following four interaction patterns.
A small group of students (red) sent only a couple of messages but wrote extremely long ones, effectively submitting a single extended response rather than engaging in a back-and-forth dialogue.
A second group (green) sent somewhat more messages, still with substantial length, suggesting engagement condensed into a handful of detailed exchanges; the majority of students (purple) fall into a middle pattern of moderate message counts with short-to-moderate message length, consistent with iterative back-and-forth engagement. 
A fourth group (cyan) sent many messages, each quite short, indicative of rapid, iterative exchanges with LLteacher. 
Across all four groups, the same inverse relationship holds: average message length drops once students exceed roughly ten messages.

\subsection{Perceived Learning Gains}

\begin{figure}[!h]
\centering
\includegraphics[width=\linewidth]{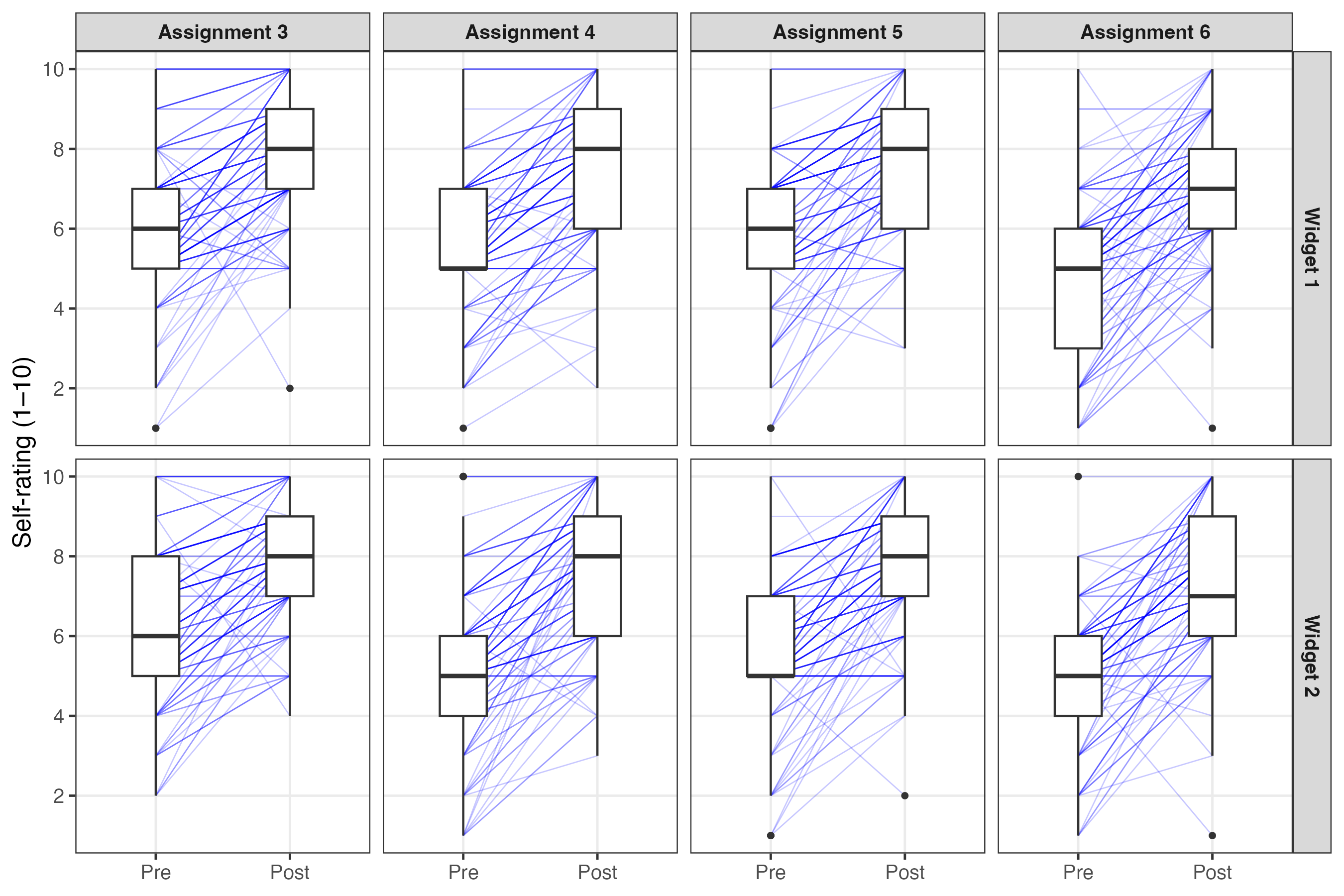}
\caption{Pre- and post-scores by assignment and widget. Lines connect individual students' scores before and after completing each LLteacher assignment. \label{fig:boxplot_learning_gains}}
\end{figure}

Students self-rated their confidence before and after each assignment on the two items described in Section~\ref{sec:data}: Widget 1 (mathematical reasoning and understanding) and Widget 2 (interpretation).
Because Assignments 1 and 2 collected these ratings through open-ended prompts rather than the 0 to 10 sliding scale, they are not considered in this analysis, but the results are available from the authors upon request.

Figure~\ref{fig:boxplot_learning_gains} shows pre- and post-scores for each widget across the four assignments which had widgets (Assignments 3--5 and the extra-credit Assignment~6). Each line traces an individual student's trajectory from pre- to post-score; boxplots summarize the distribution at each time point. 

\begin{figure}[!h]
\centering
\includegraphics[width=\linewidth]{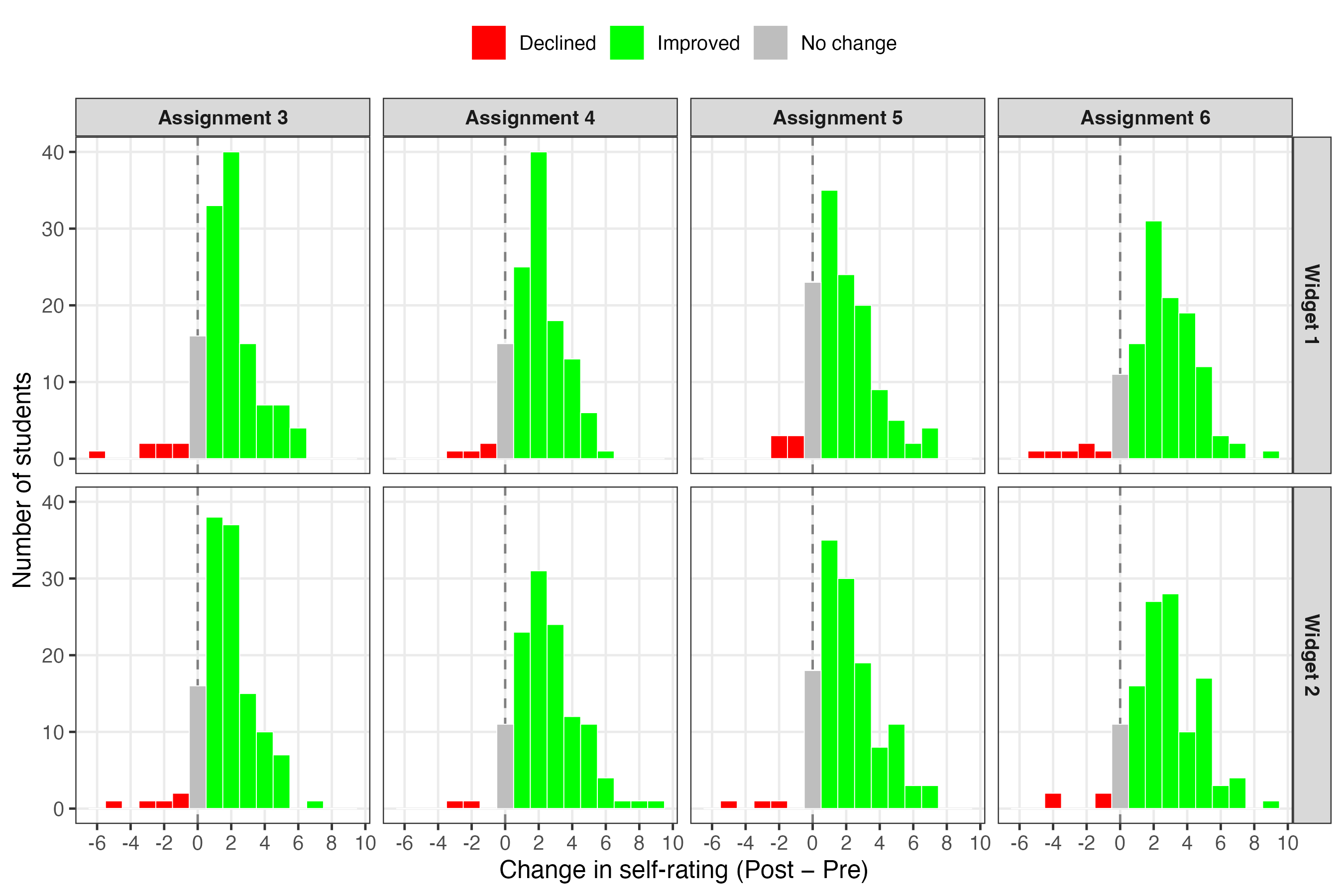}
\caption{Perceived learning gains by assignment and widget.\label{fig:hist_learning_gains}}
\end{figure}

Perceived gains are visible across all assignments and in both the interpretation and mathematical reasoning questions, with only 2.5 to 4.7\% of the students experiencing a downward trend.
Perceived learning gains, measured as post-assignment scores minus pre-assignment scores, can be seen in Figure \ref{fig:hist_learning_gains}.
As expected, the magnitude of the gains increases as the quarter progresses due to increased difficulty of the materials covered.
In addition, we note that Assignment 6 has the largest improvements and this may be because it was a discovery assignment which tackled a topic not discussed in class yet.

\begin{figure}[!h]
\centering
\includegraphics[width=\linewidth]{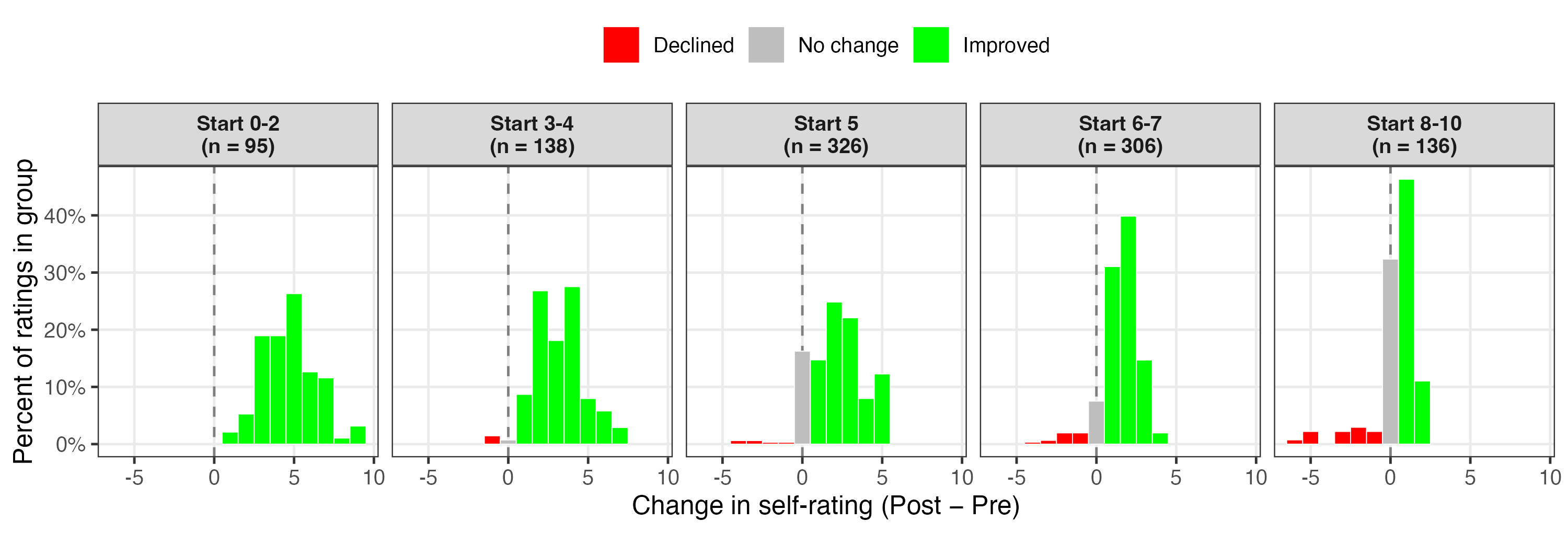}
\caption{Distribution of the change in self-rating (post minus pre) by starting-score group, pooling both widgets and Assignments 3--6. Bars show the percentage of ratings in each group, so groups of different size can be compared; $n$ is the number of ratings in the group.\label{fig:gains_by_start}}
\end{figure}

The size of the perceived gains varies quite a bit across students. What we want to make sure of is that all students benefit in the same way and, more importantly, that weaker students are being brought up toward the rest of the class.
We define weaker students as those who start an assignment with a lower self-perceived score, and we describe the gains by starting score.
Figure~\ref{fig:gains_by_start} shows the gains broken down by starting-score group.
Because some starting scores have small sample sizes, we pool the two widgets and all assignments, and we group the starting scores into five ranges, from very weak students (those who selected 0--2) to very strong students (8--10).

First, we note that the losses are mostly among students who felt stronger at the start (29 of the 37 losses come from students who started at 6 or above): it is possible that these students were overrating themselves.
More generally, as one might expect, the figure shows that the size and direction of a student's reported gain is a function of their initial score: students who started with a high score were more likely to report a smaller increase, while those who started feeling weaker were more likely to report a larger one.
This pattern likely reflects several factors: the ceiling built into the response scale (a student who starts at 10 can only stay at 10), the more general fact that there is simply less room to improve from an already-high starting point, and possibly the tool's ability to support students regardless of where they begin.

To investigate the latter while accounting for the ceiling effect, we restricted the analysis to observations with a starting value of at most 5 and a gain of at most 5, so that a 5-point gain was attainable for every observation in the sample. 
This leaves an outcome variable with only 6 values (0, 1, 2, 3, 4 and 5, we excluded losses). In order to use logistic regression, we further coded it as 1 for a large improvement (a gain of 3, 4 or 5 points) and 0 for a small or no improvement (a gain of 2 points or less).
The predictors are the starting-value group, the assignment number, and the widget type; because each student contributes up to eight observations, we include a random intercept for each student. 
Table~\ref{tab:improvement_logistic} reports the results.

\begin{table}[!h]
\centering
\small
\renewcommand{\arraystretch}{0.85}
\setlength{\tabcolsep}{5pt}
\begin{tabular}{lcc}
\toprule
Predictor & Estimate (log-odds) & Std. error \\
\midrule
Intercept & $2.759^{***}$ & 0.734 \\
\addlinespace
\multicolumn{3}{l}{\emph{Starting-value group (reference: 0--2)}} \\
\quad 3--4 & $-2.144^{***}$ & 0.622 \\
\quad 5 & $-3.719^{***}$ & 0.652 \\
\addlinespace
\multicolumn{3}{l}{\emph{Assignment (reference: Assignment 3)}} \\
\quad Assignment 4 & $-0.105$ & 0.408 \\
\quad Assignment 5 & $0.100$ & 0.428 \\
\quad Assignment 6 & $0.845^{*}$ & 0.422 \\
\addlinespace
\multicolumn{3}{l}{\emph{Widget (reference: Widget 1)}} \\
\quad Widget 2 & $0.697^{*}$ & 0.286 \\
\midrule
Observations & \multicolumn{2}{c}{512} \\
Students & \multicolumn{2}{c}{117} \\
\bottomrule
\end{tabular}
\caption{Mixed-effects logistic regression, with a random intercept for each student, of a big improvement (gain of 3 to 5 points; coded 1) versus a small or no improvement (gain of 0 to 2 points; coded 0) on starting-value group, assignment, and widget. $^{*}$ p $<$ 0.05, $^{**}$ p $<$ 0.01, $^{***}$ p $<$ 0.001.}
\label{tab:improvement_logistic}
\end{table}

Relative to students who started at 0--2, the odds of a large improvement are significantly lower for those who started at 3--4 and lower still for those who started at 5. Students were also somewhat more likely to report a large improvement on Assignment 6 and on the interpretation item (Widget 2), while Assignments 4 and 5 did not differ from Assignment 3. 
While this may seem unsurprising, this is different from a scenario in which all students gain by the same amount regardless of starting point, which would preserve the initial gap and leave students who started stronger finishing stronger as well. 
Instead, the larger gains among low-starting students suggest that we are closing the gap with their higher-starting peers, so that final outcomes converge rather than replicate the initial spread.

\subsection{Student Experience}

We evaluated student experience through an end-of-term survey. 
In particular, we were interested in three aspects: whether students found LLteacher useful for learning in the class, their satisfaction with the tool overall, and whether it changed how they relate to Statistics and to AI tools for education more broadly. 
Students rated 13 items on a five-point scale from strongly disagree to strongly agree. 

\begin{figure}[!h]
    \centering
    \includegraphics[width=\textwidth]{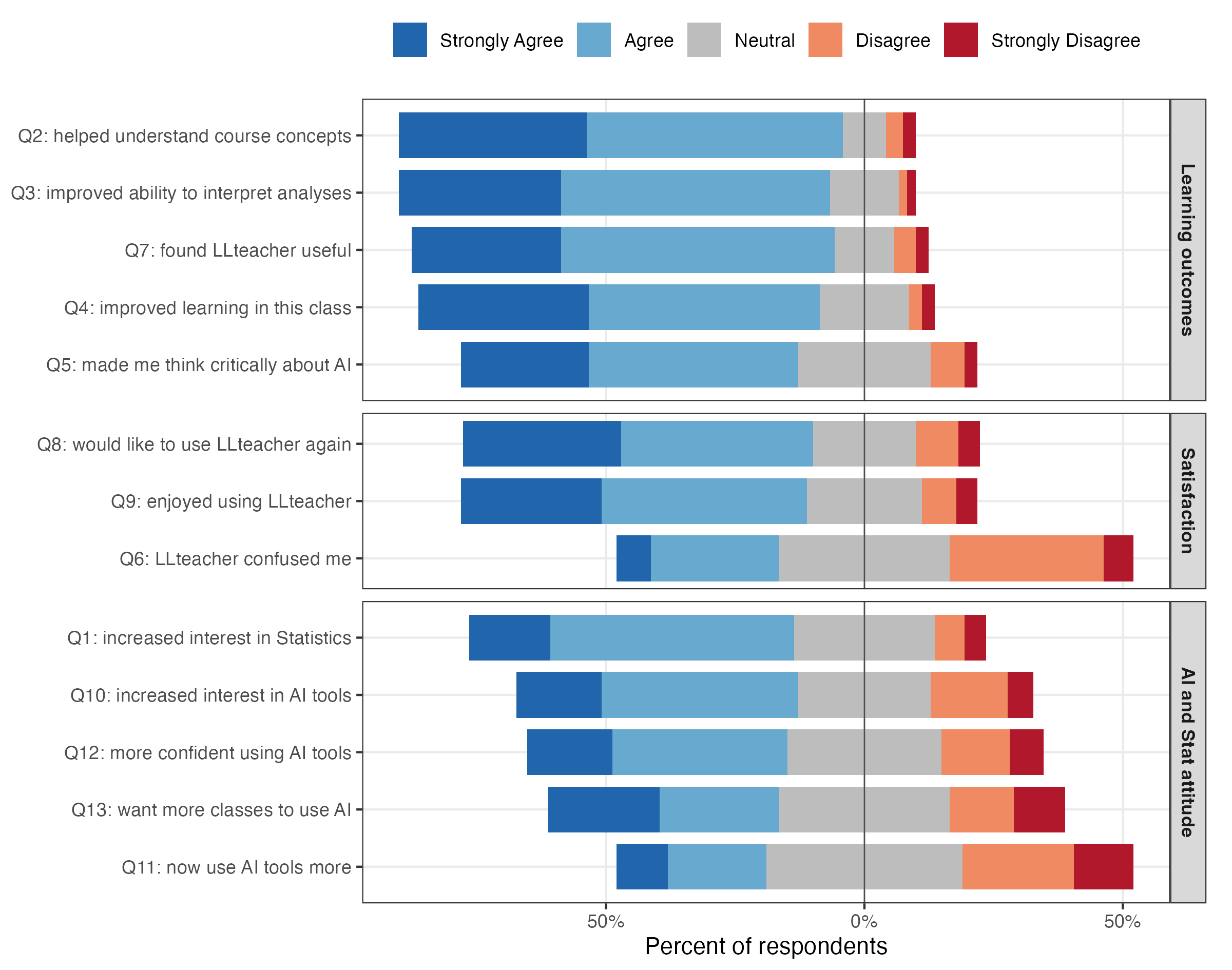}
    \caption{Distribution of student responses to the 13 rated survey items (Q1–Q13). Q6 is negatively worded, so agreement indicates a negative experience.}
    \label{fig:survey_likert_chart}
\end{figure}

The results are presented in Figure~\ref{fig:survey_likert_chart} grouped into the three dimensions we wished to evaluate: learning outcomes, satisfaction, and attitudes toward AI and Statistics.

\begin{figure}[!h]
    \centering
    \includegraphics[width=\textwidth]{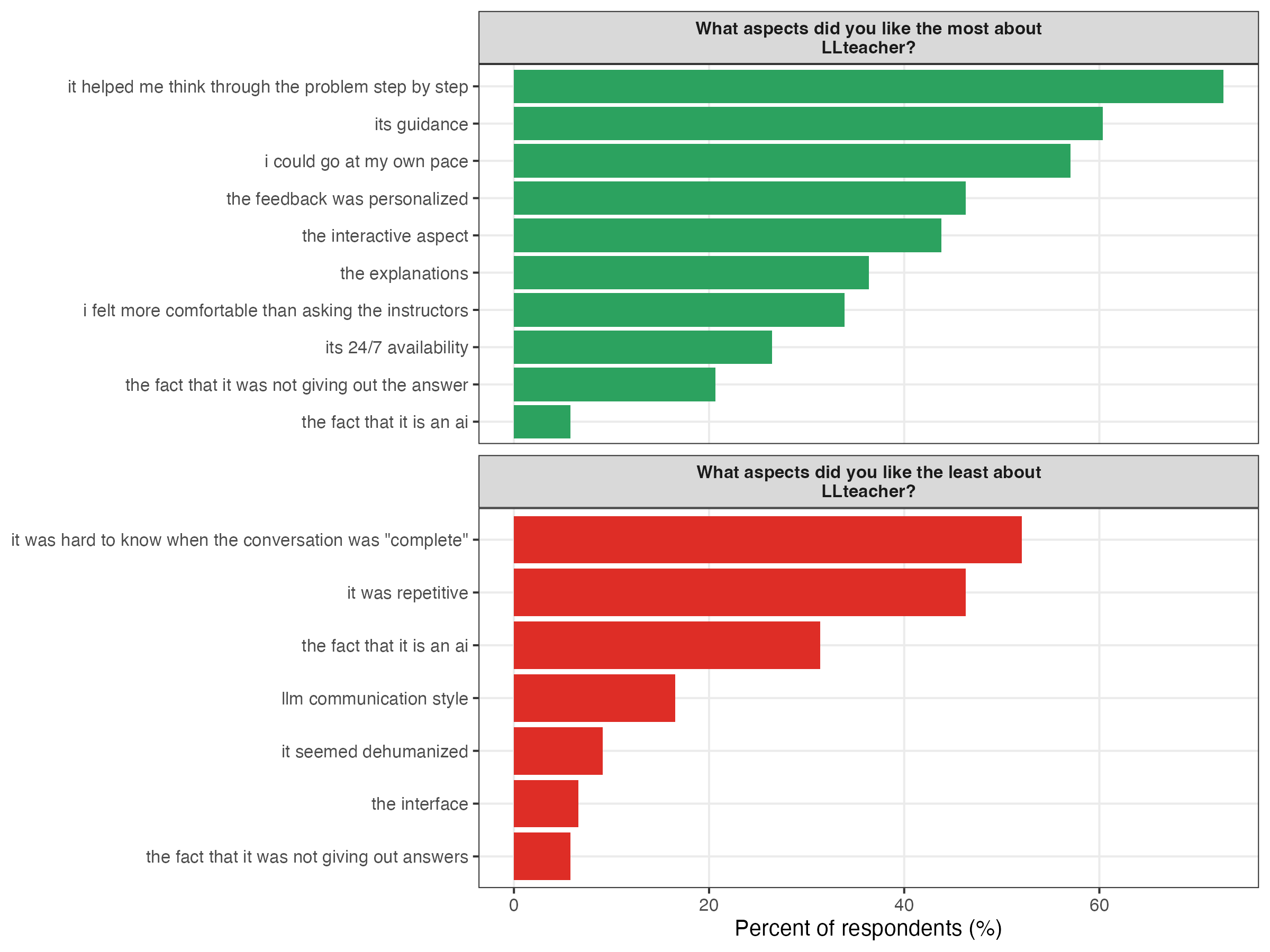}
    \caption{Aspects students liked most (Q14) and least (Q15); check-all-that-apply, so bars do not sum to 100\%.}
    \label{fig:survey_priority_chart}
\end{figure}

Overall, students responded positively to LLteacher: agreement was very high on items about understanding course concepts, interpreting analyses, and finding the tool useful (Q2, Q3, Q7), and remained high on wanting to use LLteacher again and enjoying using it (Q8, Q9).
LLteacher also appears to have increased students' interest in Statistics: 63\% agreed or strongly agreed with this statement (Q1), and only 10\% disagreed.
Responses were more mixed on whether LLteacher increased students' interest in or use of AI tools more broadly (Q10--Q13). 
One likely explanation is that many students already used AI tools regularly, including for learning, before encountering LLteacher, so a single course tool had little room to shift attitudes that were already well established.
Notably, most students disagreed that the tool confused them (Q6).

In addition to the rated items, we asked students to select, from a list, the aspects of LLteacher they liked most and least; results are shown in Figure~\ref{fig:survey_priority_chart}. 
Students most often cited that LLteacher helped them think through problems step by step, its guidance, and the ability to go at their own pace, followed by its personalized feedback and interactive nature. 
On the negative side, students most often disliked not knowing when a conversation was ``complete'', the tool's repetitive communication style, and the fact that they were interacting with an AI.

\subsubsection{Hallucinations}

\begin{figure}[!h]
    \centering
    \includegraphics[width=\textwidth]{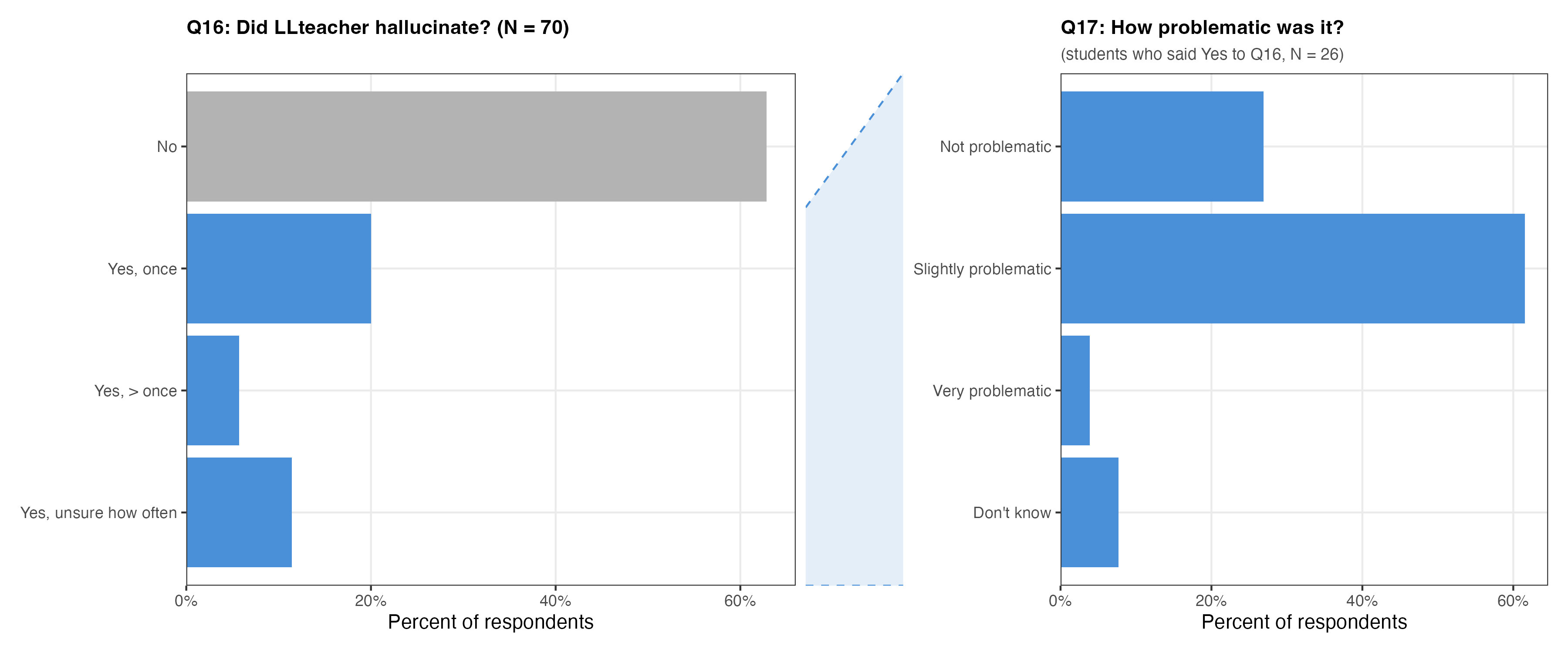}
    \caption{Percentage of students reporting whether LLteacher hallucinated (left, $N=70$) and, among those who reported a hallucination, how problematic it was for their learning or completion of the assignment (right, $N=26$).}
    \label{fig:hallucinated}
\end{figure}

Figure~\ref{fig:hallucinated} summarizes our findings regarding LLteacher's hallucinations, based on two additional survey questions: (1) ``Did LLteacher hallucinate for you?'' and (2) ``If LLteacher hallucinated, how problematic were these hallucinations for your learning or completion of the assignment?''\footnote{These questions were answered by 71 students.}.

All students who received these questions answered both, regardless of their response to the first. 
We found one inconsistent response, which was removed from this analysis.
The left panel of Figure~\ref{fig:hallucinated} shows that the majority of students (63\%) reported that LLteacher did not hallucinate responses.
We also see that 20\% said that LLteacher did hallucinate, but only once, and an additional 11\% could not recall how often the hallucinations occurred. 
Lastly, only 6\% of students reported that LLteacher hallucinated more than once.

The right panel of Figure~\ref{fig:hallucinated} shows whether students found the aforementioned hallucinations to be problematic to their learning or completion of an assignment. 
Among the students who reported a hallucination, most (62\%) found it only slightly problematic, in that it slowed their progress but they were still able to figure out the problem, and 27\% found it not problematic at all. 
Only one student reported that the hallucinations were very problematic and severely limited their learning, showing that in cases when students did encounter hallucinations, they did not find them overly problematic.

\section{Discussion}

In this paper we showed how LLteacher, a flexible LLM-based tool built for statistics education, can be used to create and administer homework assignments not merely as a supplementary tutor, but as an integral part of the class. 
LLteacher allowed us to do two things: 1) configure the LLM behavior to allow different types of assignments which would not be possible with traditional assignments and 2) see the whole conversation so that we can grade not only the final answer but the whole reasoning process of the student, where the reasoning process unfolds through a conversation with the LLM.
The study provides concrete examples of the assignments and prompts used, illustrating a practical model for integrating LLMs directly into course design and instructional activities.

We used the tool in an introductory statistics course, a course format common to many institutions, and collected data on 134 students.
Students had to use LLteacher for the six assignments of this class since these assignments would have not made sense without an LLM or with an LLM which did not follow the instructions provided.
Our results show that the vast majority of students found that LLteacher improved their understanding of the class material and their ability to interpret the results of a statistical method, and that they found the tool useful, with over 80\% of respondents agreeing or strongly agreeing with these statements. 
Results were less striking, though still positive, on items such as whether the tool made students think more critically about AI outputs. This may be because students were already inclined to think critically about AI: the course consists mostly of pre-majors who were in secondary school as AI tools became widespread and who may have already received some training on their use in prior courses. Among the attitude items, the largest shift was in students' interest in Statistics, where a majority agree or strongly agree that LLteacher increased their interest. In line with results from computer science education \citep{kazemitabaar2024codeaid, sunil-thakkar-2025, happe-et-al-2025}, these findings suggest that guided AI use, supported by scaffolding, helps students learn. The combination of increased interest in Statistics and larger confidence gains among students who started with lower confidence also suggests that LLteacher may help increase statistics accessibility.

Although the majority of the class participated in the study, participation was voluntary, and the findings are based on a single course offering during one academic quarter at a single institution, which may limit generalizability. 
The study was not designed to produce causal evidence, such as whether AI-assisted homework outperforms non-AI-assisted homework, since it did not include a control or comparison group. Similarly, the study was not designed to identify which groups of students benefit most (for example, students from underrepresented groups or with stronger academic backgrounds), so this information is not available. Learning outcomes were assessed primarily through students' self-reported confidence, measured before and after each assignment, and through self-reported perceived usefulness collected at the end of the quarter. The survey responses were not linked to students' academic records, so we do not have an objective measure of students' achievement in the course. 
This follows a substantial literature that relies on similar subjective measures both to evaluate LLM-based educational tools (see for instance \cite{kazemitabaar2024codeaid, liu2025lpitutor}) and within statistics education \citep{awuah-et-al-2023}. 
In addition, we note that the perceived learning gains observed may be temporary: we do not evaluate whether these benefits persist beyond individual assignments or translate into improved course performance and long-term retention of statistical concepts. 
Finally, although assignments were designed around recall, discovery, and critical-thinking modes, we do not have enough data to systematically examine how students responded to each type.

Future work can address these limitations in several ways. 
For instance, future studies could examine whether perceived learning gains are distributed equitably across students with different educational backgrounds, levels of preparation, and academic trajectories, by systematically collecting this information as part of the study design.
Future studies could also adopt a causal design with a treatment group to test whether LLteacher, or other tools in Statistics education, outperforms traditional homework assignments, and could be planned to collect enough data to compare how the different pedagogical modes perform; at present this is not possible, since most modes were used only once and are therefore confounded with assignment number.

Finally, using the data already collected, we plan to analyze the complete student--LLteacher conversations to better understand how learning unfolds over the course of an interaction and which interaction patterns are associated with positive outcomes; measures extracted from these conversations could also serve as an objective complement to self-reported outcomes. 
Such an analysis would also inform the design of AI tools for learning, and show how instructors can write prompts that encourage deeper student engagement.

\section*{Acknowledgments}

This deployment of LLteacher in class was supported by the ASA SSDSE Member Initiative Grant and by UW SEED AI grant.

\bibliography{bibliography}

\end{document}